\documentclass[12pt,a4wide,leqno]{article}
 \usepackage{amssymb}
 \usepackage[polish]{babel}
 \newtheorem{uw}{\hspace*{0,7cm}Uwaga}
\newtheorem{tw}{\hspace*{0,6cm}{\sc Theorem}}
\newtheorem{rem}{\hspace*{0,7cm}{\sc Remark}}
\newtheorem{lm}{\hspace*{0,6cm}{\sc Lemma}}
\newtheorem{co}{\hspace*{0,6cm}{\sc Corollary}}
\newtheorem{prop}{\hspace*{0,6cm}{\sc Proposition}}
\newtheorem{wn}{\hspace*{0,7cm}Wniosek}
\newtheorem{prz}{\hspace*{0,7cm}Przyk{\l}ad}
\newtheorem{okres}[lm]{\hspace*{0,7cm}Okre\'slenie}
\newtheorem{zad}{Zadanie}

\newcommand{\pr}{{\it Proof.}\hspace{0,2cm}}
\newtheorem{stw}[lm]{\hspace*{0,7cm}Stwierdzenie}

\newcommand{\bdf}{\begin{df}\hspace{-2mm}{\bf{.}}\hspace{2mm}}
\newcommand{\edf}{\end{df}}
\newcommand{\bokr}{\begin{okres}\hspace{-2mm}{\bf{.}}\hspace{2mm}}
\newcommand{\eokr}{\end{okres}}
\newcommand{\bzad}{\begin{zad}\hspace{-2mm}{\bf{.}}\hspace{2mm}}
\newcommand{\ezad}{\end{zad}}
\newcommand{\bstw}{\begin{stw}\hspace{-2mm}{\bf {.}}\hspace{2mm}}
\newcommand{\estw}{\end{stw}}
\newcommand{\bprz}{\begin{prz}\hspace{-2mm}{\bf{.}}\hspace{2mm}}
\newcommand{\eprz}{\end{prz}}
\newcommand{\btw}{\begin{tw}\hspace{-2mm}{\bf{.}}\hspace{2mm}}
\newcommand{\etw}{\end{tw}}
\newcommand{\btk}{\begin{tw}}
\newcommand{\etk}{\end{tw}}
\newcommand{\brem}{\begin{rem}\hspace{-2mm}{\bf{.}}\hspace{2mm}\em}
\newcommand{\erem}{\em\end{rem}}
\newcommand{\buw}{\begin{uw}\hspace{-2mm}{\bf{.}}\hspace{2mm}}
\newcommand{\euw}{\end{uw}}
\newcommand{\blm}{\begin{lm}\hspace{-2mm}{\bf{.}}\hspace{2mm}}
\newcommand{\elm}{\end{lm}}
\newcommand{\bco}{\begin{co}\hspace{-2mm}{\bf{.}}\hspace{2mm}}
\newcommand{\eco}{\end{co}}
\newcommand{\bpr}{\begin{prop}\hspace{-2mm}{\bf{.}}\hspace{2mm}}
\newcommand{\epr}{\end{prop}}
\newcommand{\bwn}{\begin{wn}\hspace{-2mm}{\bf{.}}\hspace{2mm}}
\newcommand{\ewn}{\end{wn}}

\newcommand{\B}{\hspace*{13cm}\Box}

\newcommand{\bc}{\begin{center}}
\newcommand{\ec}{\end{center}}

\begin{document}
 
\bc {\large \bf Space-time directional Lyapunov exponents}\ec 
             \bc {\large \bf for cellular automata}\ec
             \bc by \ec
             \bc {\sc M. Courbage \footnote{ Laboratoire Mati\`ere et Syst\`emes Complexes (MSC) 
UMR 7057 CNRS et Universit\'e Paris 7- Denis Diderot  
Case 7020,  Tour 24-14, 5\`eme \'etage, 
75251 Paris Cedex 05, France, 
e-mail: courbage@ccr.jussieu.fr} } and {\sc B. Kami\'{n}ski\footnote{ Faculty of Mathematics and
Computer Science, Nicolaus Copernicus University, 
Chopina 12/18, 87-100 Toru\'{n}, e-mail: bkam@mat.uni.torun.pl} } \ec

 \vspace{1cm}

 {\bf Abstract.} Space-time directional Lyapunov exponents
 are introduced. They describe the maximal velocity of propagation to the right 
or to the left of fronts of
perturbations in a  frame moving with a given velocity. The continuity
 of these exponents as function of the velocity and an inequality relating them to the
 directional entropy  is proved.\\

 \bc  {\bf 1. Introduction} \ec

In the theory of smooth dynamical
systems Lyapunov exponents describe local instability of orbits. For spatially extended systems, the 
local instability of
patterns is caused by the evolution of localised perturbations. In one dimensional
extended systems, the localised perturbations may propagate to the left or to the
right not only as travelling waves, but also as various
structures. Moreover, other phenomena, called convective instability, have been
observed in a fluid flow in a pipe, where it has been found that the system
propagates a variety of isolated and localised structures (or patches of
turbulence) moving down the pipe along the stream with some velocity.  
Convective instability has been
studied by many authors in various fields (see for example ref. 2). 
We introduce a Lyapunov
exponents describing the maximal velocity of propagation to the right 
or to the left of fronts of
perturbations in a  frame moving with a given velocity. We consider this 
problem in the framework
of one dimensional cellular automata.

 Cellular automata, first introduced by von
 Neumann have been recently used as mathematical models of natural
 phenomena $^{(1,11)}$. Lyapunov exponents in cellular
 automata have been introduced first by Wolfram $^{(11)}$. The idea
 was to find a characteristic quantity of the instability of the
 dynamics of cellular automata analogous to the Lyapunov exponents
 which measure the instability of the orbits of differentiable
 dynamical systems under perturbations of initial conditions.

 The first rigorous mathematical definition of these exponents has
 been given by Shereshevsky $^{(10)}$in the framework of ergodic
 theory.

 For a given shift-invariant probability measure $\mu$ on the
 configuration space $X$ that is also invariant under a cellular
 automaton map $f$ he defined left (resp. right) Lyapunov
 exponents $\lambda^{+}(x)$, $\lambda^{-}(x)$ as time asymptotic
 averages of the speed of the propagation to the left (resp.
 right) of a front of right (resp. left) perturbations of a given
 configuration $x\in X$. He also gave the following relation
 between the entropy and the Lyapunov exponents:
 \[h_{\mu}(f)\leq\int_{X}h_{\mu}(\sigma,x)\cdot(\lambda^{+}(x)+\lambda^{-}(x))\mu(dx)\]
 where $h_{\mu}(\sigma,x)$ denotes the local entropy of the shift
 $\sigma$ in $x$.

 In ref. 6,  another  slightly different   Lyapunov exponents have been defined for 
 cellular automata.

 But cellular automata have the richest physical and mathematical
 structure if we consider them as dynamical systems commuting with
 $\sigma$. In physical terms this reflects the local nature and the
 invariance under spatial translations of the interactions between
 cells, that is a common property of large extended systems in many
 natural applications.

 In order to account of the space-time complexity of cellular
 automata, Milnor introduced in ref. 8 a generalization of the
 dynamical entropy which he called the directional entropy. This
 concept has been later enlarged $^{(3,7)}$ to
 $\mathbb{Z}^{2}$-actions on arbitrary Lebesgue spaces.

 Here we introduce the notion of space-time directional Lyapunov
 exponents which are generalizations of the notions considered by
 Shereshevsky. We define them as the averages along a given
 space-time direction
 $\overrightarrow{v}\in\mathbb{R}\times\mathbb{R}^{+}$ of the
 speed of  propagation to the left (resp. right) of a front of
 right (resp. left) perturbations of a given configuration $x\in
 X$. We compare these exponents with the the directional entropy
 of the action generated by $\sigma$ and~$f$ (see in ref. 3, 7 and 8
 for the definition and basic properties) and we show, among other
 things, their continuity. As a corollary to our main result we
 obtain the estimation of the directional entropy given in  ref.5.

 \bc {\large\bf 2. Definitions and auxiliary results} \ec

 Let $X=S^{\mathbb{Z}}$, $S=\{0,1,\ldots,p-1\}$, $p\geq2$
 and let $\mathcal{B}$ be the $\sigma$-algebra generated by
 cylindric sets. We denote by $\sigma$ the left shift
 transformation of $X$ and by $f$ the automaton transformation
 generated by an automaton rule $F$, i.e.
 \[(\sigma x)_{i}=x_{i+1},\;\; (fx)_{i}=F(x_{i+l},\ldots,x_{i+r}),\;\;i\in\mathbb{Z},\]
 \[F:S^{r-l+1}\longrightarrow S,\;\;l,r\in\mathbb{Z},\;\;l\leq r.\]

 Let $\mu$ be a probability measure invariant w.r. to $\sigma$ and
 $f$.

 Following Shereshevsky $^{(10)}$ we put
 \[W_{s}^{+}(x)=\{y\in X;y_{i}=x_{i},i\geq s\},\]
 \[W_{s}^{-}(x)=\{y\in X;y_{i}=x_{i},i\leq-s\},\]
 $x\in X$, $s\in\mathbb{Z}$.

 It is easy to see that

 (1) The sequences $(W_{s}^{\pm}(x),s\in\mathbb{Z})$ are
 increasing, $x\in X$.

 (2) For any $a,b,c\in\mathbb{Z}$, $x\in X$ it holds
 \[\sigma^{a}W_{c}^{\pm}\left(\sigma^{b}x\right)=W_{c\mp a}^{\pm}\left(\sigma^{a+b}x\right).\]
 \blm For any $n\in\mathbb{N}$ we have
 \[f^{n}\left(W_{0}^{+}(x)\right)\subset W_{-nl}^{+}\left(f^{n}x\right),\]
 \[f^{n}\left(W_{0}^{-}(x)\right)\subset W_{nr}^{+}\left(f^{n}x\right).\]
 \elm

 \pr It is enough to show the first inclusion. Let $n=1$ and let
 $y\in W_{0}^{+}(x)$. Hence
 \[F(y_{i+l},\ldots,y_{i+r})=F(x_{i+l},\ldots,x_{i+r})\]
 for all $i\geq-l$, i.e.
 \[[f(y)]_{i}=[f(x)]_{i},\;i\geq-l,\]
 which means that $f(y)\in W^{+}_{-l}(f(x))$.

  Suppose now that \[f^{n}\left(W^{+}_{0}(x)\right)\subset W^{+}_{-nl}\left(f^{n}x\right)\]
  for some $n\in\mathbb{N}$. Using (2) one obtains
 \begin{eqnarray*}
  f^{n+1}\left(W^{+}_{0}(x)\right) & \subset & f\left(W^{+}_{-nl}\left(f^{n}x\right)\right)=f\left(\sigma^{nl}W^{+}_{0}\left(\sigma^{-nl}f^{n}x\right)\right)   \\
   & \subset & \sigma^{nl}W^{+}_{-l}\left(f\left(\sigma^{-nl}f^{n}x\right)\right)=\sigma^{nl}W^{+}_{-l}\left(\sigma^{-nl}f^{n+1}x\right)\\
   & = & \sigma^{(n+1)l}W^{+}_{0}\left(\sigma^{-(n+1)l}f^{n+1}x\right)=W^{+}_{-(n+1)l}\left(f^{n+1}x\right)
 \end{eqnarray*}
 and so the desired inequality is satisfied for any $n\in\mathbb{
 N}$.\\
 $\B$

 Lemma 1 and (1) imply at once the following\\

 {\sc Corollary.} For any $n\in\mathbb{N}$ there exists $s\in\mathbb{N}$ such
 that
 \[f^{n}\left(W^{\pm}_{0}(x)\right)\subset W^{\pm}_{s}\left(f^{n}x\right).\]

 Indeed, applying (1) it is enough to take $s=\max(0,-nl)$ in the
 case of $W^{+}(x)$ and $s=\max(0,nr)$ in the case of $W^{-}_{0}(x)$.

 Let (cf. ref.10)
 \[\widetilde{\Lambda}^{\pm}_{n}(x)=\inf\left\{s\geq0;f^{n}(W^{\pm}_{0}(x))\subset W^{\pm}_{s}\left(f^{n}x\right)\right\}\]
 and
 \[\Lambda^{\pm}_{n}(x)=\sup_{j\in\mathbb{Z}}\widetilde{\Lambda}^{\pm}_{n}\left(\sigma^{j}x\right),\]
 $x\in X$, $n\in N$.

 Obviously we have
 \[(3)\hspace*{2,4cm} 0\leq\Lambda^{+}_{n}(x)\leq\max(0,-nl),\;\;\;0\leq\Lambda^{-}_{n}(x)\leq\max(0,nr).\hspace*{2,8cm}\]

 It is shown in ref. 10, in the case $l=-r$, $r\geq0$ that the limits
 \[\lambda^{\pm}(x)=\lambda^{\pm}(f;x)=\lim_{n\rightarrow\infty}\frac{1}{n}\Lambda^{\pm}_{n}(x)\]
 exist a.e. and they are $f$ (and of course $\sigma$) - invariant
 and integrable.

 The proof given in  ref. 10 also works for arbitrary
 $l,r\in\mathbb{Z}$, $l\leq r$.

 The limit $\lambda^{+}$ (resp. $\lambda^{-}$) is called the right
 (left) Lyapunov exponent of $f$.

 It follows at once from (3) that
 \[(4)\hspace*{2,8cm} 0\leq\lambda^{+}(x)\leq\max(0,-l),\;\;\;
 0\leq\lambda^{-}(x)\leq\max(0,r).\hspace*{2,8cm}\]

 \bc {\large\bf 3. Directional Lyapunov exponents} \ec

 Let now
 $\overrightarrow{v}=(a,b)\in\mathbb{R}\times\mathbb{R}^{+}$. We
 put
 \[\alpha(t)=[at],\;\; \beta(t)=[bt],\;\; t\in\mathbb{N}\]
 where $[x]$ denotes the integer part of $x$, $x\in\mathbb{R}$.\\
 It is clear that
 \[\alpha(t_{1}+t_{2})\geq\alpha(t_{1})+\alpha(t_{2}),\;\;\;
   \beta(t_{1}+t_{2})\geq\beta(t_{1})+\beta(t_{2}),\;\; t_{1},t_{2}\in\mathbb{N}\]
 and
 \[\lim_{t\rightarrow\infty}\frac{\beta(t)}{\alpha(t)}=\frac{b}{a}, \;\; a\neq0.\]
 We put
 \[z_{l}=a+bl,\;\;\; z_{r}=a+br.\]

 \bpr For any $t\in\mathbb{N}$ we have
 \[\sigma^{\alpha(t)}f^{\beta(t)}W^{+}_{0}(x)\subset W^{+}_{-\alpha(t)-\beta(t)\cdot l}\left(\sigma^{\alpha(t)}f^{\beta(t)}x\right),\]
 \[\sigma^{\alpha(t)}f^{\beta(t)}W^{-}_{0}(x)\subset W^{-}_{\alpha(t)+\beta(t)\cdot r}\left(\sigma^{\alpha(t)}f^{\beta(t)}x\right).\]
 \epr

 \pr It follows from Lemma 1 that
 \[f^{\beta(t)}W^{+}_{0}(x)\subset W^{+}_{-\beta(t)\cdot l}\left(f^{\beta(t)}x\right).\]
 Applying (2) we get
 \[\sigma^{\alpha(t)}f^{\beta(t)}W^{+}_{0}(x)\subset\sigma^{\alpha(t)}W^{+}_{-\beta(t)\cdot l}\left(f^{\beta(t)}x\right)=
 W^{+}_{-\alpha(t)-\beta(t)\cdot l}\left(\sigma^{\alpha(t)}f^{\beta(t)}x\right).\]
 Similarly one obtains the second inclusion.\\

 {\sc Corollary.} For any $t\in\mathbb{N}$ there exists $s\in\mathbb{N}$ such
 that
 \[\sigma^{\alpha(t)}f^{\beta(t)}W^{\pm}_{0}(x)\subset W^{\pm}_{s}\left(\sigma^{\alpha(t)}f^{\beta(t)}x\right).\]

 In view of (1) it is enough to take
 \[s=\max(0,-\alpha(t)-\beta(t)\cdot l)\]
 in the case of $W^{+}_{0}(x)$ and
 \[s=\max(0,\alpha(t)+\beta(t)\cdot r)\]
 in the case of $W^{-}_{0}(x)$.

 We put
 \[\widetilde{\Lambda}^{\pm}_{\vec{v},t}(x)=\inf\left\{s\geq0;\sigma^{\alpha(t)}f^{\beta(t)}W^{\pm}_{0}(x)
 \subset W^{\pm}_{s}\left(\sigma^{\alpha(t)}f^{\beta(t)}x\right)\right\}\]
 and
 \[\Lambda^{\pm}_{\vec{v},t}(x)=\sup_{j\in\mathbb{Z}}\Lambda^{\pm}_{\vec{v},t}\left(\sigma^{j}x\right).\]

 It is clear that
 \[(5)\hspace*{3,8cm} 0\leq\Lambda^{+}_{\vec{v},t}(x)\leq\max(-\alpha(t)-\beta(t)\cdot l,0),\hspace*{4,2cm}\]
 \[0\leq\Lambda^{-}_{\vec{v},t}(x)\leq\max(\alpha(t)+\beta(t)\cdot r,0).\]

 {\sc Definition.} {\em The function $\lambda^{+}_{\vec{v}}$ (resp. $\lambda^{-}_{\vec{v}}$)
 defined by the formula
 \[\Lambda^{\pm}_{\vec{v}}(x)=\overline{\lim_{t\rightarrow\infty}}\frac{1}{t}\Lambda^{\pm}_{\vec{v},t}(x),\;\;x\in X\]
 is said to be the right (resp. left) space-time directional Lyapunov exponent of $f$.}

 We show in the sequel that in fact the limit
 $\lim_{t\rightarrow\infty}\frac{1}{t}\Lambda^{\pm}_{\vec{v},t}(x)$
 exists a.e.\\
It follows at once from (5) that
 \[(6)\hspace*{2,2cm}  0\leq\lambda^{+}_{\vec{v}}(x)\leq\max(-z_{l},0),\;\;\;
 0\leq\lambda^{-}_{\vec{v}}(x)\leq\max(z_{r},0).\hspace*{3,2cm}\]

{\bf Example}. We now consider  permutative automata. Recall that an automaton 
map $f$ defined by the rule $F$ : $S^m \rightarrow S$ is right permutative if for
 any $(\bar{x}_1,\ldots,\bar{x}_{m-l})$
the mapping: $
x_{m} \mapsto f(\bar{x}_1,\ldots,\bar{x}_{r-l},x_{m})$
is one-to-one. A left permutative mapping is defined similarly. The map $f$ is said to be
bipermutative if it is right and left permutative. 

Let $\mu$ be the uniform Bernoulli measure 
on $X$. It is well known that, due to the permutativity of $f$ (right or left), it is $f$-invariant. It is
also $\sigma$-invariant. Since the functions $\Lambda^{\pm}_{\vec{v},t} $ are $\sigma$-invariant, the
ergodicity of $\mu$ implies they are constant a.e. 

First we consider $f $ being left permutative. It follows straightforwardly from the definitions that 
$$
\Lambda^{+}_{\vec{v},t} = \max(-\alpha(t)-\beta(t)\cdot l,0)
$$
and so 
$$
\lambda^{+}_{\vec{v}} = \max(-a-b\cdot l,0) = \max(-z_{l},0).
$$
Now if  $\vec{v}$ is such that $ z_{r} = a+br <0 $ we get 
$$
\lambda^{-}_{\vec{v}}= 0.
$$
Indeed, in this case $\alpha(t)+\beta(t)\cdot r <0 $ for sufficiently large $t$, say $ t\geq t_0$ and therefore
$$
 \widetilde{\Lambda}^{-}_{\vec{v},t} = 0, t \geq t_0.
$$
Applying the continuity of the mapping $ \vec{v} \mapsto \lambda^{-}_{\vec{v}} $
 (Proposition 3) we have $ \lambda^{-}_{\vec{v}} = 0$ for $\vec{v}$ such that $ z_{r} = a+br \leq 0 $.
Similarly one checks that for $f$ being right permutative we have 
$$
\lambda^{-}_{\vec{v}}  = \max(z_{r},0)
$$
and
$$
\lambda^{+}_{\vec{v}}= 0
$$
for $\vec{v}= (a,b)$  such that $ z_{l} = a+bl \geq 0 $ . Thus if $f$ is bipermutative we have :
$$
\lambda^{+}_{\vec{v}}  = \max (-z_{l},0),  \lambda^{-}_{\vec{v}}  = \max(z_{r},0)
$$
for any $\overrightarrow{v}\in\mathbb{R}\times\mathbb{R}^{+}$.

 \blm The function $\vec{v}\longrightarrow\Lambda^{\pm}_{\vec{v}}$
 is positively homogeneous, i.e. for any $c\in\mathbb{R}^{+}$ we
 have 
 \[\lambda^{\pm}_{c\vec{v}}=c\lambda^{\pm}_{\vec{v}}.\]
 \elm

 \pr Let $(\alpha_{c}(t))$ and $(\beta_{c}(t))$ be the sequences
 associated with $c\vec{v}$, $\vec{v}=(a,b)$, i.e.
 $\alpha_{c}(t)=~[cat]~=~\alpha(ct)$, $\beta_{c}(t)=[cbt]=\beta(ct)$.
 It is clear that
 \[\Lambda^{\pm}_{c\vec{v},t}(x)=\Lambda^{\pm}_{\vec{v},ct}(x),\;\;x\in X\]
 which implies at once the desired equality.

 \blm For any
 $\vec{v}\in\mathbb{R}\times\mathbb{R}^{+},\;s,t\in\mathbb{N}$ and
 $x\in X$ we have
 \[\Lambda^{\pm}_{\vec{v},s+t}(x)\leq\Lambda^{\pm}_{\vec{v},s}(x)+\Lambda^{\pm}_{\vec{v},t}\left(f^{\beta(s)}x\right)+2|l|.\]
 \elm

 \pr We shall consider only the case of
 $\Lambda^{+}_{\vec{v},s+t}$; the proof for
 $\Lambda^{-}_{\vec{v},s+t}$ is similar, $s,t\geq0$.

 We put
 \[\widetilde{s}=\Lambda^{+}_{\vec{v},s}(x),\;\;\;\widetilde{t}=\Lambda^{+}_{\vec{v},t}\left(f^{\beta(s)}x\right).\]

 By the definition and the $\sigma$-invariance of
 $\Lambda^{+}_{\vec{v},t}$we have
 \begin{eqnarray*}
  (7)\hspace{1,5cm}
  \sigma^{\alpha(s)+\alpha(t)}f^{\beta(s)+\beta(t)}W^{+}_{0}(x)&
  \subset &
  \sigma^{\alpha(t)}f^{\beta(t)}W^{+}_{\widetilde{s}}\left(\sigma^{\alpha(s)}f^{\beta(s)}x\right)\\
  & = &
  \sigma^{\alpha(t)}f^{\beta(t)}\sigma^{-\widetilde{s}}W^{+}_{0}\left(\sigma^{\widetilde{s}+\alpha(s)}f^{\beta(s)}x\right)\\
  & \subset &
  \sigma^{-\widetilde{s}}W^{+}_{\widetilde{\Lambda}^{+}_{\vec{v},t}(\sigma^{\widetilde{s}+\alpha(s)}f^{\beta(s)}x)}\left(\sigma^{\alpha(t)}f^{\beta(t)}\sigma^{\widetilde{s}+\alpha(s)}f^{\beta(s)}x\right)\\
  & \subset &
  \sigma^{-\widetilde{s}}W^{+}_{\widetilde{t}}\left(\sigma^{\widetilde{s}}\sigma^{\alpha(s)+\alpha(t)}f^{\beta(s)+\beta(t)}x\right)\\
  & = &
  \sigma^{-\widetilde{s}-\widetilde{t}}W^{+}_{0}\left(\sigma^{\widetilde{s}+\widetilde{t}}\sigma^{\alpha(s)+\alpha(t)}f^{\beta(s)+\beta(t)}x\right)\\
  & = &
  W^{+}_{\widetilde{s}+\widetilde{t}}\left(\sigma^{\alpha(s)+\alpha(t)}f^{\beta(s)+\beta(t)}x\right).\\
 \end{eqnarray*}

 We put
 \[\delta_{\alpha}=\alpha(s+t)-\left(\alpha(s)+\alpha(t)\right),\;\;\;\delta_{\beta}=\beta(s+t)-(\beta(s)+\beta(t)),\]
 $s,t\geq0$.

 Acting on the both sides of (7) by
 $\sigma^{\delta_{\alpha}}f^{\delta_{\beta}}$ we obtain by Lemma 1
 \begin{eqnarray*}
  (8)\hspace{2,5cm}
  \sigma^{\alpha(s+t)}f^{\beta(s+t)}W^{+}_{0}(x)
  & \subset &
  \sigma^{\delta_{\alpha}}f^{\delta_{\beta}}W^{+}_{\widetilde{s}+\widetilde{t}}\left(\sigma^{\alpha(s)+\alpha(t)}f^{\beta(s)+\beta(t)}x\right)\\
  & = &
  f^{\delta_{\beta}}W^{+}_{\widetilde{s}+\widetilde{t}-\delta_{\alpha}}\left(\sigma^{\alpha(s+t)}f^{\beta(s)+\beta(t)}x\right)\\
  & \subset &
  W^{+}_{\widetilde{s}+\widetilde{t}-\delta_{\alpha}-\delta_{\beta\cdot l}}\left(\sigma^{\alpha(s+t)}f^{\beta(s+t)}x\right).\hspace*{2,5cm}\\
 \end{eqnarray*}

 Since $0\leq\delta_{\alpha}$, $\delta_{\beta}\leq2$ we get from (8)
 \begin{eqnarray*}
 \widetilde{\Lambda}^{+}_{\vec{v},s+t}(x) & \leq &
 \max\left(\widetilde{s}+\widetilde{t}-\delta_{\alpha}-\delta_{\beta}l,0\right)\\
 & \leq & \widetilde{s}+\widetilde{t}+2|l|\\
 & = &
 \Lambda^{+}_{\vec{v},s}(x)+\Lambda^{+}_{\vec{v},t}\left(f^{\beta(s)}x\right)+2|l|
 \end{eqnarray*}
 which implies at once the desired inequality.\\
 $\B$

 \bpr For any $\vec{v}\in\mathbb{R}\times\mathbb{R}^{+}$ and
 almost all $x\in X$ there exists the limit
 \[\lim_{t\rightarrow\infty}\frac{1}{t}\Lambda^{\pm}_{\vec{v},t}(x)=\lambda^{\pm}_{\vec{v}}(x).\]
 \epr

 \pr It is enough to consider the case of $\Lambda^{+}_{\vec{v},t}, \;t\geq0.$

 First we consider the case $\vec{v}=(a,1)$, i.e.
 \[\alpha(t)=[at],\;\;\;\beta(t)=t,\;\;t>0.\]

 In this case Lemma 3 has the form
 \[\Lambda^{\pm}_{\vec{v},s+t}(x)\leq\Lambda^{\pm}_{\vec{v},s}(x)+\Lambda^{\pm}_{\vec{v},t}\left(f^{s}x\right)+2|l|,\]
 $s,t\geq0.$ It is easy to see that this inequality permits to
 apply the Kingman subadditive ergodic theorem, which implies the
 existence a.e. of the limit
 \[\lim_{t\rightarrow\infty}\frac{1}{t}\Lambda^{\pm}_{\vec{v},t}(x)=\lambda^{\pm}_{\vec{v}}(x).\]

 The case of arbitrary $\vec{v}$ easily reduces to the above by
 Lemma 2.\\
 $\B$

 \bpr The space-time directional Lyapunov exponents
 $\lambda^{\pm}_{\vec{v}}$,
 are continuous as functions of 
$\vec{v}$ for $\vec{v}\in\mathbb{R}\times\left(\mathbb{R}^{+}\setminus\{0\}\right)$ .
 \epr

 \pr The result will easily follow from the inequality
 \[(9)\hspace*{3,5cm} b\cdot\lambda^{+}_{\vec{v}'}(x)\leq\max\left(b'\lambda^{+}_{\vec{v}}(x)-(a'b-ab'),0\right)\hspace{4cm}\]
 where $\vec{v}=(a,b),\;\vec{v}'=(a',b'),\;b,b'>0$.

 In order to show (9) we first consider the case $b'=b=1$. In this
 case $\beta(t)~=~\beta'(t)~=~t,\;t>0$. We put
 $p(t)=\alpha'(t)-\alpha(t)$, $t>0$.

 We have
 \begin{eqnarray*}
 \sigma^{\alpha'(t)}f^{\beta'(t)}W^{+}_{0}(x) & = & \sigma^{\alpha(t)+p(t)}f^{\beta(t)}W^{+}_{0}(x)\\
 & = & \sigma^{p(t)}\sigma^{\alpha(t)}f^{\beta(t)}W^{+}_{0}(x)\\
 & \subset & \sigma^{p(t)}W^{+}_{\widetilde{\Lambda}^{+}_{\vec{v},t}(x)}\left(\sigma^{\alpha(t)}f^{\beta(t)}x\right)\\
 & = & W^{+}_{\widetilde{\Lambda}^{+}_{\vec{v},t}(x)-p(t)}\left(\sigma^{p(t)}\sigma^{\alpha(t)}f^{\beta(t)}x\right)\\
 & = & W^{+}_{\widetilde{\Lambda}^{+}_{\vec{v},t}(x)-p(t)}\left(\sigma^{\alpha'(t)}f^{\beta'(t)}x\right)\\
 & \subset & W^{+}_{\Lambda^{+}_{\vec{v},t}(x)-p(t)}\left(\sigma^{\alpha'(t)}f^{\beta'(t)}x\right)\\
 & \subset &
 W^{+}_{\max(\Lambda^{+}_{\vec{v},t}(x)-p(t),0)}\left(\sigma^{\alpha'(t)}f^{\beta'(t)}x\right)
 \end{eqnarray*}
 and so
 \[\widetilde{\Lambda}^{+}_{\vec{v}',t}(x)\leq\max\left(\Lambda^{+}_{\vec{v},t}(x)-p(t),0\right)\]
 which implies
 \[(10) \hspace*{4,5cm} \lambda^{+}_{\vec{v}'}(x)\leq\max\left(\lambda^{+}_{\vec{v}}(x)-(a'-a),0\right),\hspace*{5,1cm}\]
 i.e (9) is satisfied for $b=b'=1$.

 Let now $\vec{v},\vec{v}'$ be arbitrary, $b',b>0$. We have
 \[\vec{v}=b\cdot\vec{v}_{0},\;\;\vec{v}'=b'\cdot\vec{v}'_{0}\]
 where
 \[\vec{v}_{0}=\left(\frac{a}{b},1\right),\;\;\vec{v}'_{0}=\left(\frac{a'}{b'},1\right).\]
 It follows from (10) that
 \[\lambda^{+}_{\vec{v}'_{0}}(x)\leq\max\left(\lambda^{+}_{\vec{v}_{0}}(x)-\left(\frac{a'}{b'}-\frac{a}{b}\right),0\right)\]
 and therefore, by the use of the homogeneity of
 $\lambda^{+}_{\vec{v}}$, we obtain
 \[\frac{1}{b'}\lambda^{+}_{\vec{v}'}(x)\leq\max\left(\frac{1}{b}\lambda^{+}_{\vec{v}}(x)-\left(\frac{a'}{b'}-\frac{a}{b}\right),0\right)\]
 which gives (9) for arbitrary $\vec{v},\vec{v}'$, $b,b'>0$.

 Let now $\vec{v}=(a,b),\;b>0$ be fixed and let
 $\vec{v}'_{n}=(a'_{n},b'_{n})$ be such that
 $\vec{v}'_{n}\longrightarrow\vec{v}$ as $n\longrightarrow\infty$.

 It follows from (9) and (10), respectively, that
 \[\overline{\lim_{n\rightarrow\infty}}\lambda^{+}_{\vec{v}'_{n}}(x)\leq\max\left(\lambda^{+}_{\vec{v}}(x),0\right)=\lambda^{+}_{\vec{v}}(x),\]
 \[\lambda^{+}_{\vec{v}}(x)\leq\max\left(\lim_{\overline{n\rightarrow\infty}}\lambda^{+}_{\vec{v}'_{n}}(x),0\right)=\lim_{\overline{n\rightarrow\infty}}\lambda^{+}_{\vec{v}'_{n}}(x)\]
 i.e.
 \[\lim_{n\rightarrow\infty}\lambda^{+}_{\vec{v}'_{n}}(x)=\lambda^{+}_{\vec{v}}(x)\]
 which gives the desired result.

 Let now $\Phi$ be the action generated by $\sigma$ and $f$ and let
 $h^{\mu}_{\vec{v}}(\Phi)$ denote the directional entropy of
 $\Phi$ in the direction $\vec{v}$.

 {\sc Theorem.} {\em For any $\vec{v}=(a,b)$, $a\in\mathbb{R},\;b\geq0$ and any
 $\Phi$-invariant measure $\mu$ we have
 \[h^{\mu}_{\vec{v}}(\Phi)\leq
\int_{X}h_{\mu}(\sigma,x)\left(\lambda^{+}_{\vec{v}}(x)+\lambda^{-}_{\vec{v}}(x)\right)\mu(dx)\]
where $h_{\mu}(\sigma,x)$ is the local entropy of $\sigma$ at the point $x  ^{(10)}$. 
In particular, if $\mu $ is ergodic with repect to $\sigma$, then $ \lambda^{\pm}_{\vec{v}}$ are constant a.e. and }

$$
h^{\mu}_{\vec{v}}(\Phi)\leq h_{\mu}(\sigma)\left(\lambda^{+}_{\vec{v}}+\lambda^{-}_{\vec{v}}\right).
$$
\pr First we consider the case
 $\vec{v}=(p,q)\in\mathbb{Z}\times\mathbb{N}$. In this case it is easy to show that
 \[h^{\mu}_{\vec{v}}(\Phi)=h_{\mu}(\sigma^{p}f^{q}),\;\;\lambda^{+}_{\vec{v}}(x)=\lambda^{+}(\sigma^{p}f^{q};x).\]
 Since $\sigma^{p}f^{q}$ is an automaton map Theorem of  ref. 10
 implies
 \begin{eqnarray*}
 (11)\hspace*{2cm} h^{\mu}_{\vec{v}}(\Phi) & = & h_{\mu}(\sigma^{p}f^{q})\leq\int_{X}h_{\mu}(\sigma,x)
 \left(\lambda^{+}\left(\sigma^{p}f^{q};x\right)+\lambda^{-}\left(\sigma^{p}f^{q};x\right)\right)\mu(dx)\\
 & = &
 \int_{X}h_{\mu}(\sigma,x)\left(\lambda^{+}_{\vec{v}}(x)+\lambda^{-}_{\vec{v}}(x)\right)\mu(dx).\hspace{4cm}
 \end{eqnarray*}
 The homogeneity of the mappings
 \[\vec{v}\longrightarrow h^{\mu}_{\vec{v}}(\Phi),\;\;\vec{v}\longrightarrow\lambda^{+}_{\vec{v}}\]
 and (11) imply that the inequality
 \[(12)\hspace*{3,2cm}h^{\mu}_{\vec{v}}(\Phi)\leq\int_{X}h_{\mu}(\sigma,x)\left(\lambda^{+}_{\vec{v}}(x)+\lambda^{-}_{\vec{v}}(x)\right)\mu(dx)\hspace*{3,8cm}\]
 is valid for every $\vec{v}\in\mathbb{Q}\times\mathbb{Q}^{+}$.

 Let now $\overrightarrow{v}=(a,b)\in\mathbb{R}\times\mathbb{R}^{+}$.
 If $b=0$ then the desired inequality is satisfied because for
 $\vec{v}=(1,0)$ we have $h_{\vec{v}}(\Phi)=h_{\mu}(\sigma)$, $\lambda^{+}_{\vec{v}}(x)=0,\;\lambda^{-}_{\vec{v}}(x)=1,\;x\in
 X$. Thus let us suppose $b>0$ and let
 $\vec{v}_{n}=(a_{n},b_{n})\longrightarrow\vec{v}$.
 Hence and from the inequalities
 \[0\leq\lambda^{+}_{\vec{v}_{n}}(x)\leq\max(0,-z^{(n)}_{l}),\;\;0\leq\lambda^{-}_{\vec{v}_{n}}(x)\leq\max(0,-z^{(n)}_{r})\]
 where $z^{(n)}_{l}=a_{n}+b_{n}l$, $z^{(n)}_{r}=a_{n}+b_{n}r$,
 $n\geq1$ it follows that the sequence
 $\left(\lambda^{+}_{\vec{v}_{n}}(x)+\lambda^{-}_{\vec{v}_{n}}(x)\right)$ is
 jointly bounded. By Proposition 3
 \[\lambda^{\pm}_{\vec{v}_{n}}(x)\longrightarrow\lambda^{\pm}_{\vec{v}}(x)\;\;\;\textrm{a.e.}\]

 It follows from ref.9 that the mapping
 $\vec{v}\longrightarrow h^{\mu}_{\vec{v}}(\Phi)$ is continuous. It
 is well known (cf. ref. 4) that the function
 $x\longrightarrow h_{\mu}(\sigma,x)$ is integrable.

 Therefore applying the Lebesgue dominated convergence theorem we
 get from (12) the desired inequality for all
 $\vec{v}\in\mathbb{R}\times\mathbb{R}^{+}$. The inequality in the ergodic case follows at once from the
Brin-Katok formula:
$$
\int_{X}h_{\mu}(\sigma,x)\mu(dx) = h_{\mu}(\sigma)
$$
\\
 $\B$

 It follows at once from the above theorem

 \bco For any $\vec{v}\in\mathbb{R}\times\mathbb{R}^{+}$ and for
 any $\Phi$-invariant measure $\mu$ we have
 \[h^{\mu}_{\vec{v}}(\Phi)\leq h_{\mu}(\sigma)\left(\max(0,-z_{l})+\max(0,z_{r})\right).\]
 \eco

 It is interesting to see that our computation of the Lyapounov exponents in the example of section 3 and the
results of ref.5 ( Theorems 1 and 2) imply the following :\\

{\bf Remark}.
The inequality given in the above theorem (and Corollary 1) becomes the equality in the following cases:\\
i)   $f$ left permutative  and $z_r \leq 0$,\\
ii)  $f $ right permutative  and $z_l \geq 0$,\\
iii) $ f$ left permutative  and $z_l \leq 0$ , $z_l \geq 0$. \\

 From Corollary 1 one easily obtains the following
estimation for
 $h^{\mu}_{\vec{v}}(\Phi)$ proved by us in ref.5.

 \bco For any $\vec{v}\in\mathbb{R}\times\mathbb{R}^{+}$ and for
 any $\Phi$-invariant measure $\mu$ we have
 \[h^{\mu}_{\vec{v}}(\Phi)\leq\max\left(|z_{l}|,|z_{r}|\right)\log p\;\;\;\ if \;\;\; z_{l}\cdot z_{r}\geq0\]
 and
 \[h^{\mu}_{\vec{v}}(\Phi)\leq|z_{r}-z_{l}|\log p\;\;\;\ if \;\;\; z_{l}\cdot z_{r}\leq0.\]
 \eco

 \bc {\large\bf References} \ec
 $1. $ J.P.Allouche, M.Courbage, J.P.S.Kung, G.Skordev,
 {\em Cellular automata, Encyclopedia of Physical Science and
 Technology}, Third Edition, vol. 2, 2002, 555-567, Academic
 Press.\\
$2$. T. Bohr, D.A. Rand, 
 {\em A mechnism for localised turbulence}, Physica D, 52 (1991), 532-543
 $3$. M.Boyle, D.Lind, {\em Expansive subdynamics}, Trans. Amer.
 Math. Soc. 349 (1997), 55-102.\\
 $4$. M.Brin, A.Katok, {\em On local entropy}, Lecture Notes in
 Math. 470, Springer Verlag, New York, 30-38.\\
 $5$. M.Courbage, B.Kami{\'n}ski, {\em On the directional entropy of
$\mathbb{Z}^{2}$-actions
 generated by cellular automata}, Studia Math. 153 (2002),
 285-295.\\
 $6$.P. Tisseur {\em Cellular automata and Lyapunov exponents}, Nonlinearity 13 (2000), 1547-1560.\\
 $7$. B.Kami{\'n}ski, K.K.Park, {\em On the directional entropy of
$\mathbb{Z}^{2}$-action on a Lebesque
 space}, Studia Math. 133 (1999), 39-51.\\
 $8$. J.Milnor, {\em On the entropy geometry of cellular automata},
 Complex Systems 2 (1988), 357-386.\\
 $9$. K.K.Park, {\em On directional entropy functions}, Israel J.
 Math. 113 (1999), 243-267.\\
 $10$. M.A.Shereshevsky, {\em Lyapunov exponents for one-dimensional
 automata}, J. Nonlinear Sci. 2 (1992), 1-8.\\
 $11$. S.Wolfram, {\em Cellular automata and complexity},
 Addison-Wesley Publishing Company 1994.\\
 
 \end{document}